\documentclass[a4paper]{PoS}

\usepackage[retainorgcmds]{IEEEtrantools}
\usepackage{amsmath}
\usepackage{fixltx2e} 
\usepackage[hide]{ed}

\usepackage{mytikzpackage} 

\newcommand{\pdiff}[2]{\frac{\partial #1}{\partial #2}}

\title{Improving Polynomial-filtered Hybrid Monte Carlo With Hasenbusch}

\ShortTitle{Improving Polynomial-filtered Hybrid Monte Carlo With Hasenbusch}

\author{\speaker{Taylor Haar}\\
        CSSM, Department of Physics, The University of Adelaide, Adelaide, SA, Australia 5005\\
        E-mail: \email{taylor.haar@adelaide.edu.au}}
     
\author{Waseem Kamleh\\
        CSSM, Department of Physics, The University of Adelaide, Adelaide, SA, Australia 5005\\
        E-mail: \email{waseem.kamleh@adelaide.edu.au}}

\author{James Zanotti\\
        CSSM, Department of Physics, The University of Adelaide, Adelaide, SA, Australia 5005\\
        E-mail: \email{james.zanotti@adelaide.edu.au}}

\author{Yoshifumi Nakamura\\
        RIKEN Advanced Institute for Computational Science, Kobe, Hyogo 650-0047, Japan\\
        E-mail: \email{nakamura@riken.jp}}

\abstract{
The predominant method for generating Lattice QCD configurations is Hybrid Monte Carlo (HMC). In order to speed up this generation,
a wide range of preconditioning techniques that modify the lattice action have been devised.
This work compares the performance of the well-known Hasenbusch preconditioning technique with the polynomial filtering technique on a small $16^3 \times 32$ lattice with two flavours of Wilson fermions at a pion mass $M_{\pi} \sim 400$~MeV.
We explore a novel method of combining polynomial and Hasenbusch filters, revealing a speedup when compared to the standard two Hasenbusch filters. This comes with the added advantage of simplified tuning.
}

\FullConference{The 26th International Nuclear Physics Conference\\
		11-16 September, 2016\\
		Adelaide, Australia}

\begin{document}

\section{Introduction and Motivation} \label{sec:introduction}
Lattice QCD is the non-perturbative method of choice when dealing with strong interactions.
In order to measure observables on the lattice, we numerically evaluate path integrals via an ensemble average over a large number of configurations, each of which are described by the state of the gauge field $U$ and fermion field $\psi$.
In order to generate these configurations, we most commonly use Hybrid Monte Carlo (HMC), which involves repeated inversions of the Dirac matrix $M$ that describes the strong force between fermions at any two lattice sites.
The large number of inversions required and the size of the matrix involved mean that HMC is very computational intensive, making it difficult to simulate near physical quark masses.

A large variety of optimization techniques have been applied to HMC.
These techniques reduce the number of matrix inversions required or improve the condition number of the fermion matrix such that the overall cost is reduced.
In our work \cite{Haar:2016bwe}, we examine mass preconditioning \cite{Hasenbusch:2003} and polynomial filtering \cite{Kamleh:2011dc} on a $16^3 \times 32$ lattice to compare their performance, before investigating a more novel technique where these two methods are combined.
In section \ref{sec:theory}, we explain the formulation of both methods before the initial comparison in section \ref{sec:pf_v_mp}. Then in section \ref{sec:pf_w_mp} we investigate the performance benefits of using polynomial filters on top of mass preconditioners.

\section{Theory} \label{sec:theory}
The standard method for generating lattice configurations with dynamical fermions is Hybrid Monte Carlo (HMC).
The starting point is the lattice action
\begin{equation}
	S = S_G[U] + S_F[U, \psi, \overline{\psi}],
\end{equation}
consisting of a gluonic part $S_G$ that depends purely on the gauge fields $U$ and a fermionic part $S_F$ that also depends on the fermion fields $\psi, \overline{\psi}$.
We wish to find configurations $(U, \psi, \overline{\psi})$ distributed according to $\exp(-S[U, \psi, \overline{\psi}])$.
To do this, we first use Wick's theorem to convert the fermions field $\psi$ to bosonic pseudofermion fields $\phi$ to make them computationally friendly, modifying the fermion action in the process.
We then need to sample $\phi$ from the distribution $\exp [-S_F(U, \phi, \phi^\dag)]$, which can be done in practise by relating to Gaussian noise vectors $\chi$ distributed according to $\exp [-\chi^\dag \chi]$.
To generate correctly distributed gauge fields $U$, we introduce a fictitious conjugate momentum field $P$ distributed according to $\exp [-\sum \mathrm{tr}[P^2]]$, then produce configurations which preserve the Hamiltonian
\begin{equation}
	H[U,\phi,\phi^\dag] = \sum \mathrm{tr}[P^2] + S[U,\phi, \phi^\dag]
\end{equation}
by using Hamilton's equations, leading to integration steps
\begin{IEEEeqnarray}{rrcl}
	&\hat{T}[\epsilon]&:& (P,U) \rightarrow  (P, e^{i\epsilon P} U) \IEEEyesnumber \IEEEyessubnumber \label{eq:time_step} \\
\mathrm{and}\quad &\hat{S}[\epsilon]&:& (P,U) \rightarrow  (P - \epsilon F, U), \IEEEyessubnumber \label{eq:space_step}
\end{IEEEeqnarray}
where $F = \pdiff{S}{U}$ is the force term.
We combine a sequence of these steps into a trajectory to take an existing configuration $(P, U)$ and produce a new state $(P', U')$, such that we get to a new candidate configuration $(U', \phi')$.
This candidate undergoes a Metropolis acceptance step to ensure detailed balance and hence that the process will converge to the target distribution $\exp[-S]$.

The main cost arises from the fermionic force term, which in the two-flavour degenerate case $S_F = \phi^\dag (M^\dag M)^{-1} \phi$ takes the form
\begin{equation}
	\pdiff{S_F}{U} = \pdiff{}{U}[\phi^\dag (M^\dag M)^{-1} \phi] = - \phi^\dag (M^\dag M)^{-1} \pdiff{M^\dag M}{U} (M^\dag M)^{-1} \phi.
\end{equation}
The issue here is the inversion of the fermion matrix $K \equiv M^\dag M$. 
It is a very large, sparse matrix, so it takes iterative solvers (e.g.\ conjugate gradient) many matrix multiplications to invert.
It can also give rise to large force terms, which then necessitates a reduction in the step-size to keep a good Metropolis acceptance rate and thus increases the number of required inversions.
Therefore, modern lattice simulations use a variety of HMC improvements  to reduce the force terms (thus reducing the number of required inversions) or to make the matrix $K$ easier to invert.

A template for achieving a reduction in computational cost was proposed by \cite{Peardon:2002wb}, which considers splitting the action into two terms
\begin{equation}
	S = S_{UV} + S_{IR}, \label{eq:multiterm}
\end{equation}
such that
$S_{UV}$ captures the high-energy modes ($\sim$ large forces) of the action whilst $S_{IR}$ captures the low-energy modes ($\sim$ small forces), and
$F_{UV}$ is relatively cheap to calculate compared to $F_{IR}$.
Such terms can then be placed on different time-scales using a multiple  time-scale integrator, which allows the expensive $F_{IR}$ to be evaluated less often and hence improve the cost.

One of the standard HMC improvement techniques is mass preconditioning (MP) \cite{Hasenbusch:2003}, where we factorize the fermion action $S_F$ into two terms as follows:
\begin{equation}
	S_{MP} = \phi_1^\dag J^{-1} \phi_1  + \phi_2^\dag JK^{-1} \phi_2.
\end{equation}
Here, $J$ is a fermion matrix just like $K$, but with a modified mass parameter $\kappa' < \kappa$ giving rise to a `heavier' quark mass.
The first action term $S_1$ captures the high-energy modes but has a cheaper force term than $S_2$, so we can use multiple time-scales to reduce the overall cost.

We compare this technique to polynomial filtering (PF) \cite{Kamleh:2011dc}, whereby the fermion action is filtered via
\begin{equation}
	S_{PF} = \phi_1^\dag P(K) \phi_1 + \phi_2^\dag [P(K)K]^{-1},
\end{equation}
where $P(K)$ is a low-order polynomial approximating $K^{-1}$. In our work, we use Chebyshev polynomials such that the only parameter to tune is the polynomial order $p$. The polynomial filter term $S_1$ has a very cheap force term due to a lack of matrix inversions, and it captures the high-energy modes of the system by virtue of $P(K)$ approximating the inverse. Hence, we can put $S_1$ and $S_2$ on separate time-scales to achieve a cost reduction.

\section{Comparison of filtering methods} \label{sec:pf_v_mp}
\subsection{Setup}
Our initial study compares the computational cost of polynomial filtering against mass preconditioning.
We use a $n_f = 2$, $16^3 \times 32$ lattice with a Wilson fermion action with hopping parameter $\kappa = 0.15825$, and even-odd preconditioning. This lattice has a pion mass of $\sim 400\ \mathrm{MeV}$ and a lattice spacing of $\sim 0.08\ \mathrm{fm}$.

The metric we use for the cost is
\begin{equation}
	C = \frac{N_{mat}}{P_{acc}} \label{eq:cost}
\end{equation}
where $N_{mat}$ is the number of fermion matrix $K$ multiplications per trajectory and $P_{acc}$ is the Metropolis acceptance rate. We average this quantity over at least 2000 trajectories for each displayed data point in the graphs that follow.

As noted in equation (\ref{eq:multiterm}), we split the different action terms onto different integration time-scales. By way of example, the single polynomial filter action (1PF) takes the form
\begin{equation}
	S = S_G + \phi_1^\dag P(K) \phi_1 + \phi_2^\dag [P(K)K]^{-1} \phi_2,
\end{equation}
and we set the step-sizes $h_0 = h_G < h_1 < h_2$ to reflect the relative cost and frequency scales.
The corresponding number of steps $n_i$ at each scale are given by the relation $\tau = h_i n_i$ where $\tau$ is the trajectory length; we have fixed $\tau = 1$ as is standard.
We use a generalized integration scheme (see section \ref{sec:genint}) that gives a very flexible choice of step-sizes, and the `balanced forces' method to tune these step-sizes. See the paper \cite{Haar:2016bwe} for further details.

\subsection{Generalized multi-scale integration} \label{sec:genint}
To create multiple time-scales in a HMC integration, one typically uses a nested integration scheme. Finer integration scales are built by substituting integration schemes for the time updates in the original integration scheme.
For example, consider the space-time-space leapfrog integrator
\begin{equation}
	I_{LPF}[\tau] = \left( \hat{S}\left[\tfrac{h}{2}\right] \hat{T}\left[ h \right] \hat{S}\left[ \tfrac{h}{2} \right] \right)^{n},
\end{equation}
where $n = \tau/h$ and $\hat{T}, \hat{S}$ are as defined in \eqref{eq:time_step}, \eqref{eq:space_step}. To add a nested integration scale to this scheme, we can replace the time updates $\hat{T}[h]$ with $m$ leapfrog steps in the second term $S_2$:
\begin{IEEEeqnarray}{rCl}
	I_2[h] & = & \left( \hat{S}_2 \left[\tfrac{h}{2}\right] \hat{T}[h] \hat{S}_2\left[\tfrac{h}{2}\right] \right)^{m}, \\
	\mathrm{s.t.} \quad I_{\mathrm{nested}}[\tau] & = & \left( \hat{S}_1 \left[ \tfrac{h_1}{2} \right] I_2[h_1] \hat{S}_1\left[ \tfrac{h_1}{2} \right] \right)^{n}.
\end{IEEEeqnarray}
This ensures that $S_1$ is integrated with step-size $h_1 = \tau/n$ and $S_2$ is integrated with finer step-size $h_2 = h_1/m$.
However, nested schemes force the step-sizes of each scale to evenly divide those on each coarser scale.

We have devised an generalized scheme for constructing multiple integration time-scales without any relative step-size restriction.
The basic idea is to note that when we integrate some Hamiltonian $H = T + S = T + S_1 + S_2 + \ldots$, we always have just one kind of time step $\hat{T}[h]$ that is applied in a uniform direction $h > 0$.
It thus makes sense to parametrize the progress of time-steps via a time parameter $t$ that increases monotonically from $0$ to $\tau$.
We can then combine integration schemes for Hamiltonians for each action term $H_i = T + S_i$ into a generalized scheme for the full Hamiltonian $H = T + S$ by integrating with time steps from $0$ to $\tau$, inserting the action step updates $\hat{S}_i[h]$ at their respective `times' $t$ in the composite integrators.
For example, consider a two-step and a three-step leapfrog integrator for the two action terms:
\begin{IEEEeqnarray}{rCl}
	I_1[\tau] & = & \hat{S}_1 \left[ \tfrac{\tau}{4} \right] \hat{T} \left[ \tfrac{\tau}{2} \right] \hat{S}_1 \left[ \tfrac{\tau}{2} \right]  \hat{T} \left[ \tfrac{\tau}{2} \right] \hat{S}_1 \left[ \tfrac{\tau}{4} \right], \IEEEyesnumber \IEEEyessubnumber \\
	\mathrm{and} \quad I_2[\tau] & = & \hat{S}_2 \left[ \tfrac{\tau}{6} \right] \hat{T} \left[ \tfrac{\tau}{3} \right] \hat{S}_2 \left[ \tfrac{\tau}{3} \right]  \hat{T} \left[ \tfrac{\tau}{3} \right] \hat{S}_2 \left[ \tfrac{\tau}{3} \right] \hat{T} \left[ \tfrac{\tau}{3} \right] \hat{S}_2 \left[ \tfrac{\tau}{6} \right]. \IEEEyessubnumber
\end{IEEEeqnarray}
We can combine these two schemes via the generalized scheme by overlaying the space updates based on their position in `time'. This produces the integrator
\begin{equation}
	I[\tau] =
\hat{S}_1 \left[ \tfrac{\tau}{4} \right]
 \hat{S}_2 \left[ \tfrac{\tau}{6} \right]
\hat{T} \left[ \tfrac{\tau}{3} \right]
 \hat{S}_2 \left[ \tfrac{\tau}{3} \right]
\hat{T} \left[ \tfrac{\tau}{6} \right]
\hat{S}_1 \left[ \tfrac{\tau}{2} \right]
\hat{T} \left[ \tfrac{\tau}{6} \right]
 \hat{S}_2 \left[ \tfrac{\tau}{3} \right]
\hat{T} \left[ \tfrac{\tau}{3} \right]
 \hat{S}_2 \left[ \tfrac{\tau}{6} \right]
\hat{S}_1 \left[ \tfrac{\tau}{4} \right]
\end{equation}
Further details are given in \cite{Haar:2016bwe}, where we also prove that this method produces an integration scheme that meets the requirements for detailed balance, given that the composite integrators also do so.

\subsection{Results}
\begin{figure}
\centering
\begin{tikzpicture}[baseline, trim left=(group c1r1.south west)]

\begin{groupplot}[
	group style={
		group size=2 by 1,
		horizontal sep=0pt,
		yticklabels at=edge left,
		},
	ymin=0, ymax=120000,
	small,
	minor y tick num=3,
	scaled y ticks=base 10:-4,
	y errors,
	ymajorgrids,
	xlabel style={
		at={(0.5,-0.3)}, 
		anchor=mid,
	},
	]

\nextgroupplot[
	title={Polynomial},
	xlabel={$p$},
	ylabel={Cost},
	only marks,
	xtick=data,
	cycle list name=mstone_d,
	enlarge x limits=0.25,
]

\renewcommand{\matrixopfile}{\figdir/data/witers_j131X.txt}



\addplot+[red, mark=*]
	table[
		x=p,
		y expr={\thisrow{iter_SF} + \thisrow{iter_F1} + \thisrow{iter_F2}},
		y error expr={\thisrow{iter_tot_err}*\autocorr},
	]
	{\matrixopfile};

\nextgroupplot[
	title={Mass prec.\makebox[0pt]{\phantom{y}}}, 
	xlabel={$\kappa'$},
	only marks,
	xtick={0.154, 0.1545, 0.155, 0.1555, 0.156},
	cycle list name=mstone_d,
	xticklabel style={
		rotate=45,
		anchor=east,
		/pgf/number format/precision=4,
	},
]


\renewcommand{\matrixopfile}{\figdir/data/witers_j133X.txt}
	
	
\addplot+[red, mark=*]
	table[
		x=rho,
		y expr={\thisrow{iter_SF} + \thisrow{iter_F1} + \thisrow{iter_F2}},
		y error expr={\thisrow{iter_tot_err}*\autocorr},
	]
	{\matrixopfile};

\end{groupplot}
\end{tikzpicture}
\caption{The simulation cost function \eqref{eq:cost} for (single filter) polynomial filtering and mass preconditioning.} \label{fig:1f_cost}
\end{figure}
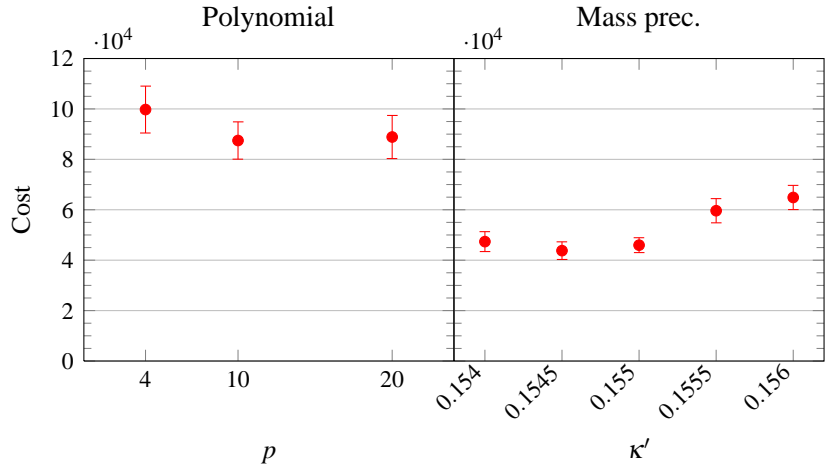

Figure \ref{fig:1f_cost} shows the cost function $C$ \eqref{eq:cost} for a single polynomial filter and for a single mass preconditioner.
As one can see, the mass preconditioned action performs much better: a cost of $C = 43,800 \pm 3,500$ at $\kappa' = 0.1545$ compared with $C = 87,500 \pm 7,400$ at $p = 10$.
This suggests that a short order polynomial of order $10$ can't capture as much of the dynamics as a mass preconditioner that requires $80$ or more iterations to invert.

We can improve the performance of both methods by using two filters instead of one. In the case of polynomial filtering, we take two polynomials $P_1(K)$ and $P_2(K)$ with orders $p_2 > p_1$ that factorize into a polynomial $Q(K) = P_2(K)/P_1(K)$, such that we can use action
\begin{equation}
	S_{2PF} = \phi_1^\dag P_1(K) \phi_1 + \phi_2^\dag Q(K) \phi_2  + \phi_3^\dag [P_2(K)K]^{-1} \phi_3 
\end{equation}
and thus keep the cost of evaluating the intermediate force $F_2$ low.
In the case of mass preconditioning, we use two modified mass parameters $\kappa_1 < \kappa_2 < \kappa$, and action
\begin{equation}
	S_{2MF} = \phi_1^\dag J_1^{-1} \phi_1 + \phi_2^\dag J_1 J_2^{-1} \phi_2  + \phi_3^\dag J_2 K^{-1} \phi_3.
\end{equation}
To keep the parameter space manageable, we fixed the cheapest term $S_1$ in both cases for our analysis -- namely, $p_1 = 4$ and $\kappa_1 = 0.145$.

\begin{figure}
\centering
\begin{tikzpicture}[baseline, trim left=(group c1r1.south west)]

\begin{groupplot}[
	group style={
		group size=2 by 1,
		horizontal sep=0pt,
		yticklabels at=edge left,
		},
	ymin=0, ymax=70000,
	small,
	minor y tick num=4,
	ymajorgrids,
	y errors,
	xlabel style={
		at={(0.5,-0.3)}, 
		anchor=mid,
	},
	]

\nextgroupplot[
	title={2PF},
	xlabel={$p_2$},
	ylabel={Cost},
	only marks,
	xtick=data,
	xticklabels={$24$,$34$,$54$},
	cycle list name=mstone_d,
	enlarge x limits=0.25,
]

\renewcommand{\matrixopfile}{\figdir/data/witers_j132X.txt}


\addplot+[red, mark=*]
	table[
		x=q,
		y expr={\thisrow{iter_SF} + \thisrow{iter_F1} + \thisrow{iter_F2} + \thisrow{iter_F3}},
		y error expr={\thisrow{iter_tot_err}*\autocorr},
	]
	{\matrixopfile};

\nextgroupplot[
	title={2MP},
	xlabel={$\kappa_2$},
	only marks,
	cycle list name=mstone_d,
	xticklabel style={
		rotate=45,
		anchor=east,
		/pgf/number format/precision=4,
	},
]

\renewcommand{\matrixopfile}{\figdir/data/witers_j134X.txt}


\addplot+[red, mark=*]
	table[
		x=rho2,
		y expr={\thisrow{iter_SF} + \thisrow{iter_F1} + \thisrow{iter_F2} + \thisrow{iter_F3}},
		y error expr={\thisrow{iter_tot_err}*\autocorr},
	]
	{\matrixopfile};

\end{groupplot}
\end{tikzpicture}
\caption{The cost function \eqref{eq:cost} for 2 filter actions, comparing polynomial filtering (2PF) and mass-preconditioning (2MP).} \label{fig:2f_cost}
\end{figure}
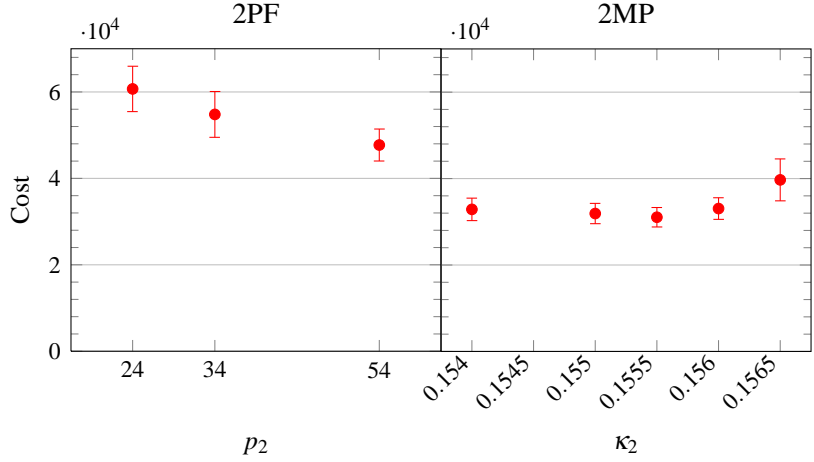

Figure \ref{fig:2f_cost} shows the cost of these two methods side by side.
In both cases we see improved results over the one-filter case. 
Two polynomial filters reduce the cost approximately as much as a single mass preconditioner: a cost of $C = 47,700 \pm 3,700$ at $p_2 = 54$.
Two mass preconditioners reduce the cost further, to $C = 31,100 \pm 2,200$ at $\kappa_2 = 0.1555$.

\section{Combined filters} \label{sec:pf_w_mp}
\subsection{Combining polynomial filtering and mass-preconditioning}
It is informative to compare the two techniques' relative efficiency.
Polynomial filtering works better at capturing the high-frequency (energy) modes of the system, because we can then use a low polynomial order. To capture lower frequency modes, we have to increase the polynomial order and the cost to evaluate the force terms increases dramatically.
On the other hand, Hasenbusch filtering is best suited to small changes $\Delta \kappa = \kappa' - \kappa$ in the quark mass, because this implies that $J K^{-1} \approx I$ and hence the correction term is easier to invert. However, the filter term $\phi_1^\dag J^{-1} \phi_1$ would then have a more expensive force term due to the lighter mass.

To combine the benefits of both of these methods, we place a polynomial filter on top of a mass preconditioner, resulting in the action
\begin{equation}
	S_{\mathrm{PF-MP}} = \phi_1^\dag P(J) \phi_1 + \phi_2^\dag [P(J)J]^{-1}\phi_2 + \phi_3^\dag JK^{-1} \phi_3
\end{equation}
where $J = K(\kappa'), \kappa' > \kappa$ and $P(J)$ is a polynomial.
The idea here is that the high frequency modes are captured by the cheap polynomial $P(J)$, such that the relative mass of the Hasenbusch correction term $\Delta \kappa$ can be kept small.

We tested this on our lattice with the polynomial $P(K)$ fixed to order $p=4$ and $\kappa'$ varying. The cost function for various $\kappa'$ is shown in Figure \ref{fig:pfmp_cost} in the second column, which can be compared to the 2MP action in the first column.
As the graph shows, the cost for PF-MP is very similar to that of 2MP.

\subsection{3-filter actions}
Note that in the cost function for 2MP and PF-MP, the cost increases significantly depending on the choice of the intermediate term parameter $\kappa_2 / \kappa'$, so a degree of tuning is required to achieve a minimum.
This is not a cheap procedure, as we need at least 500 trajectories to get a decent handle on the acceptance rate $P_{acc}$ for each set of parameters and hence the cost.
This is complicated by the fact that the Hasenbusch parameter $\kappa_2 / \kappa'$ is a real parameter whose optimal value is heavily dependent on the quark mass $\kappa$, so we'd have to tune it again for different quark masses.
This makes it difficult to optimize an action with three mass preconditioners effectively, because there are too many parameters $\kappa_1, \kappa_2, \kappa_3$ which require fine tuning.

On the other hand, polynomial filtering only depends on a single integral parameter $p$ once a class of polynomials is chosen, and the same polynomial order should filter out similar proportions of the action regardless of $\kappa$.
Hence, a polynomial filtering term requires very little to no tuning once a good set of polynomials are found.
We can thus add more polynomial filters to our action without increasing the time it takes to tune.

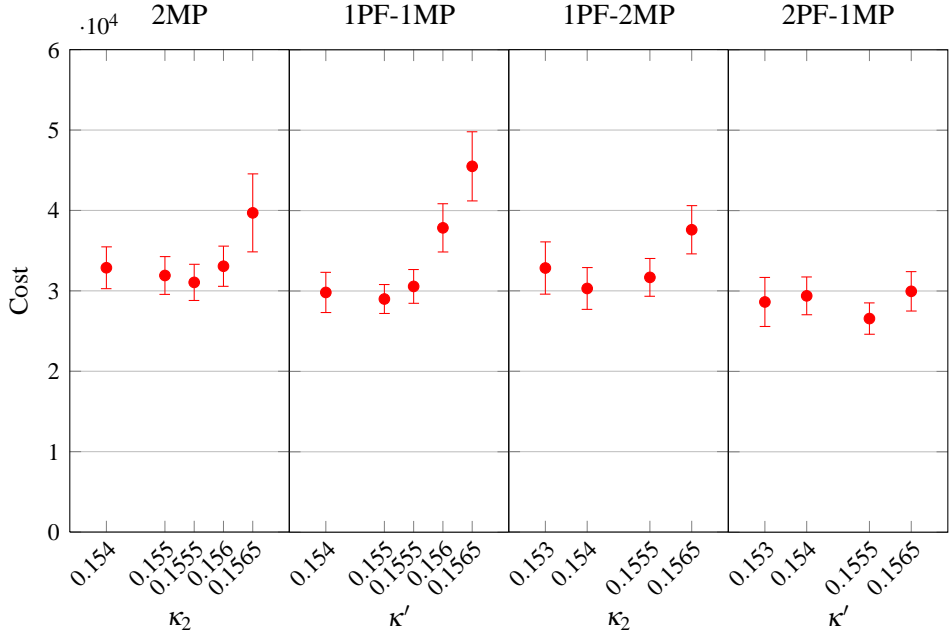
\begin{figure}
\centering


\renewcommand{\autocorr}{5} 

\begin{tikzpicture}[baseline, trim left=(group c1r1.south west)]

\begin{groupplot}[
	group style={
		group size=4 by 1,
		horizontal sep=0pt,
		yticklabels at=edge left,
		},
	ymin=0, ymax=60000,
	small,
	width=4.5cm,
	height=8cm,
	scaled y ticks=base 10:-4,
	ytick scale label code/.code={},
	xtick=data,
	y errors,
	xlabel style={
		at={(0.5,-0.18)},
		anchor=mid,
	},
	ymajorgrids=true,
	]
	
\nextgroupplot[
	title={2MP},
	xlabel={$\kappa_2$},
	ylabel={Cost},
	only marks,
	cycle list name=mstone_d,
	slanted xlabels,
	enlarge x limits=0.25,
	ytick scale label code/.code={$\cdot 10^{#1}$},
]

\renewcommand{\matrixopfile}{\figdir/data/witers_j134X.txt}

\addplot+[red, mark=*]
	table[
		x=rho2,
		y expr={\thisrow{iter_SF} + \thisrow{iter_F1} + \thisrow{iter_F2} + \thisrow{iter_F3}},
		y error expr={\thisrow{iter_tot_err}*\autocorr},
	]
	{\matrixopfile};

\nextgroupplot[
	title={1PF-1MP},
	xlabel={$\kappa'$},
	only marks,
	xtick=data,
	cycle list name=mstone_d,
	slanted xlabels,
	enlarge x limits=0.25,
]

\renewcommand{\matrixopfile}{\figdir/data/witers_j137X1.txt}

\addplot+[red, mark=*]
	table[
		x=rho,
		y expr={\thisrow{iter_SF} + \thisrow{iter_F1} + \thisrow{iter_F2} + \thisrow{iter_F3}},
		y error expr={\thisrow{iter_tot_err}*\autocorr},
	]
	{\matrixopfile};

\nextgroupplot[
	title={1PF-2MP},
	xlabel={$\kappa_2$},
	only marks,
	cycle list name=mstone_d,
	slanted xlabels,
	enlarge x limits=0.25,
]

\renewcommand{\matrixopfile}{\figdir/data/witers_j1301X.txt}

\addplot+[red, mark=*]
	table[
		x=rho2,
		y expr={\thisrow{iter_SF} + \thisrow{iter_F1}
		+ \thisrow{iter_F2} + \thisrow{iter_F3}
		+ \thisrow{iter_F4}},
		y error expr={\thisrow{iter_tot_err}*\autocorr},
	]
	{\matrixopfile};

\nextgroupplot[
	title={2PF-1MP},
	xlabel={$\kappa'$},
	only marks,
	cycle list name=mstone_d,
	slanted xlabels,
	enlarge x limits=0.25,
]

\renewcommand{\matrixopfile}{\figdir/data/witers_j1302X.txt}

\addplot+[red, mark=*]
	table[
		x=rho,
		y expr={\thisrow{iter_SF} + \thisrow{iter_F1}
		+ \thisrow{iter_F2} + \thisrow{iter_F3}
		+ \thisrow{iter_F4}},
		y error expr={\thisrow{iter_tot_err}*\autocorr},
	]
	{\matrixopfile};

\end{groupplot}
\end{tikzpicture}
\caption{The cost function \eqref{eq:cost}, comparing mass preconditioning (2MP) with polynomial filtered mass preconditioning (1PF-1MP, 2PF-1MP, 1PF-2MP)} \label{fig:pfmp_cost}
\end{figure}

We consider one polynomial filter on top of two mass preconditioners (1PF-2MP)
\begin{equation}
	S_{1PF-2MP} = \phi_1^\dag P(J_1) \phi_1 + \phi_2^\dag [P(J_1) J_1]^{-1} \phi_2 + \phi_3^\dag J_1 J_2^{-1} \phi_3 + \phi_4^\dag J_2, K^{-1} \phi_4
\end{equation}
where we fix $p=4$, $\kappa_1 = 0.145$; and two polynomial filters on top of a single mass preconditioner (2PF-1MP)
\begin{equation}
	S_{2PF-1MP} = \phi_1^\dag P(J) \phi_1 + + \phi_2^\dag Q(J) \phi_2 + \phi_3^\dag [P_2(J) J]^{-1} \phi_3 + \phi_4^\dag J K^{-1} \phi_4,
\end{equation}
where we fix $p=4$, $q=20$.
The cost function for these two actions are shown in columns three and four respectively in Figure \ref{fig:pfmp_cost}.

As can be seen in Figure \ref{fig:pfmp_cost}, the optimal cost for the two three-filter actions are very similar to those for PF-MP and 2MP.
The extra polynomial filter in 1PF-2MP on top of 2MP has no effect on the performance.
However, for 2PF-1MP the cost has a much lower dependence on the choice of the Hasenbusch parameter $\kappa$, as can be seen at the $\kappa' = 0.1565$ data point.
Since the polynomial filter orders $(p,q)$ don't require much tuning, this indicates that the 2PF-1MP action doesn't require much tuning to reach close-to-optimum computational costs.
In fact, it requires even less tuning than 2MP.

\section{Conclusion and Outlook} \label{sec:conclusion}
We have compared the performance of polynomial filtering and mass preconditioning on a modest $16^3 \times 32$ lattice before combining the two techniques to try to achieve lower costs.
Whilst we didn't observe any significant improvements in cost over the `standard' 2MP algorithm, we did find that 2PF-1MP cost much less to tune due to the cost's low dependence on $\kappa'$ and the fact that the polynomial filters don't require very much tuning.
It remains to be seen if this behaviour still holds for larger lattices and for smaller quark masses.

\bibliographystyle{JHEP}
\bibliography{references}

\providecommand{\href}[2]{#2}\begingroup\raggedright\begin{thebibliography}{1}

\bibitem{Haar:2016bwe}
T.~Haar, W.~Kamleh, J.~Zanotti and Y.~Nakamura, \emph{{Applying polynomial
  filtering to mass preconditioned Hybrid Monte Carlo}},
  \href{https://arxiv.org/abs/1609.02652}{{\tt 1609.02652}}.

\bibitem{Hasenbusch:2003}
M.~Hasenbusch and K.~Jansen, \emph{{Speeding up the HMC: QCD with clover
  improved Wilson fermions}},
  \href{http://dx.doi.org/10.1016/S0920-5632(03)01737-7}{\emph{Nucl.Phys.Proc.Suppl.}
  {\bf 119} (2003) 982--984},
  [\href{https://arxiv.org/abs/hep-lat/0210036}{{\tt hep-lat/0210036}}].

\bibitem{Kamleh:2011dc}
W.~Kamleh and M.~Peardon, \emph{{Polynomial Filtered HMC: An Algorithm for
  lattice QCD with dynamical quarks}},
  \href{http://dx.doi.org/10.1016/j.cpc.2012.05.002}{\emph{Comput.Phys.Commun.}
  {\bf 183} (2012) 1993--2000}, [\href{https://arxiv.org/abs/1106.5625}{{\tt
  1106.5625}}].

\bibitem{Peardon:2002wb}
{\scshape TrinLat Collaboration} collaboration, M.~J. Peardon and J.~Sexton,
  \emph{{Multiple molecular dynamics time scales in hybrid Monte Carlo fermion
  simulations}},
  \href{http://dx.doi.org/10.1016/S0920-5632(03)01738-9}{\emph{Nucl.Phys.Proc.Suppl.}
  {\bf 119} (2003) 985--987},
  [\href{https://arxiv.org/abs/hep-lat/0209037}{{\tt hep-lat/0209037}}].

\end{thebibliography}\endgroup

\end{document}